\documentclass[sigconf]{acmart}

\copyrightyear{2024}
\acmYear{2024}
\setcopyright{acmlicensed}\acmConference[CHI '24]{Proceedings of the CHI Conference on Human Factors in Computing Systems}{May 11--16, 2024}{Honolulu, HI, USA}
\acmBooktitle{Proceedings of the CHI Conference on Human Factors in Computing Systems (CHI '24), May 11--16, 2024, Honolulu, HI, USA}
\acmDOI{10.1145/3613904.3642916}
\acmISBN{979-8-4007-0330-0/24/05}

\begin{document}
\title[Exploring Patients' Challenges and Tool Needs for Self-Management of Postoperative Acute Pain]{``It's Sink or Swim'': Exploring Patients' Challenges and Tool Needs for Self-Management of Postoperative Acute Pain}

\author{Souleima Zghab}
\email{souleima.zghab@polymtl.ca}
\affiliation{%
  \department{Department of Computer and Software Engineering}
  \institution{Polytechnique Montreal}  \city{Montreal}
  \state{QC}
  \country{Canada}
}

\author{Gabrielle Pag\'e}
\email{gabrielle.page@umontreal.ca}
\affiliation{%
  \department{Department of Anesthesiology and Pain Medicine}
  \institution{University of Montreal}
  \city{Montreal}
  \state{QC}
  \country{Canada}
}

\author{M\'elanie Lussier}
\email{melanie_lussier@hotmail.fr}
\affiliation{%
  \institution{Patient Author}
  \city{Montreal}
  \state{QC}
  \country{Canada}
}

\author{Sylvain B\'edard}
\email{sylvain.bedard@ceppp.ca}
\affiliation{%
  \institution{Patient Author}
  \city{Montreal}
  \state{QC}
  \country{Canada}
}

\author{Jinghui Cheng}
\email{jinghui.cheng@polymtl.ca}
\affiliation{%
  \department{Department of Computer and Software Engineering}
  \institution{Polytechnique Montreal}
  \city{Montreal}
  \state{QC}
  \country{Canada}
}

\begin{abstract}
    Poorly managed postoperative acute pain can have long-lasting negative impacts and pose a major healthcare issue. There is limited investigation to understand and address the unique needs of patients experiencing acute pain. In this paper, we tackle this gap through an interview study with 14 patients who recently underwent postoperative acute pain to understand their challenges in pain self-management and their need for supportive tools. Our analysis identified various factors associated with the major aspects of acute pain self-management. Together, our findings indicated that tools for supporting these patients need to carefully consider information and support delivery to adapt to rapid changes in pain experiences, offer personalized and dynamic assistance that adapts to individual situations in context, and monitor emotion when promoting motivation. Overall, our work provided valuable knowledge to address the less-investigated but highly-needed problem of designing technology for the self-management of acute pain and similar health conditions.
\end{abstract}



\maketitle

\section{Introduction}
The inadequate management of pain is widely recognized as a major healthcare issue~\cite{Campbell2021, rathmell2006acute}, with over 80\% of patients experiencing moderate-to-severe acute pain after a surgical procedure requiring an incision~\cite{apfelbaum2003postoperative}. Poorly managed postoperative acute pain is associated with a broad range of negative consequences, including impaired function, delayed recovery, development of chronic pain, and prolonged use of opioid-based analgesia~\cite{Katz2012,gan2017poorly}. The self-management of acute pain is multifaceted, including not only the management of analgesic intake, but also adopting nonpharmacologic strategies, such as cooling and compression, breathing, and relaxation, to control pain in the patient's daily life. Post-discharge patients are often on their own when managing pain via these pharmacological, physical, and psychological strategies \cite{nimmo2017enhanced}. Ideally, these strategies should ``focus on the patient's own efforts in controlling pain rather than depending on external agents'', such as medication~\cite{zhu2017postoperative}. But in reality, the easier access to pharmacological treatments compared to physical and psychological strategies often results in patients relying more heavily on medications for pain relief~\cite{Wunsch2016, Page2020}. To be successful, patients need to make judgments about their level of pain and analgesic use. Their self-management skills can be influenced by their knowledge, behavior, and attitude toward their health condition, including pain~\cite{jahn2014improvement}. So when it comes to strategies to manage postoperative pain, it is crucial to address the patient's mental state -- their beliefs about pain and the way those beliefs affect their behaviors.

New technologies may help make pain self-management more effective, accessible, and affordable~\cite{10.1145/3453892.3461350,10.1145/3491101.3503562}. However, the current exploration of technologies for pain management mostly focuses on chronic pain (e.g.,~\cite{Gromala2015, Singh2017, 10.1145/3025453.3025832}). Management of acute pain, however, has several distinctive characteristics that pose unique challenges in technology design and thus require attention from the HCI community. Acute pain often originates from rare incidences (e.g., injury or surgery) that develop quickly and are unfamiliar to patients. Patients are thus usually under-prepared and experience a variety of emotions such as anxiety and stress with the new and uncertain experience of pain~\cite{Walker2016}. Moreover, poorly managed acute pain can have long-lasting negative impacts, including the development of chronic pain~\cite{Katz2012}. For a minority of patients, the initial temporary use of analgesics can also continue for months after the surgery~\cite{Page2020}. Additionally, the design of technologies needs to consider the relatively short time frame of acute pain and the altered cognitive and emotional states patients are in, as a result of postoperative recovery and medication intake. The success of digital tools thus relies on their ability to motivate and engage patients, establishing trust while providing the needed support, within a short period of time. Because of these unique characteristics, it is crucial to investigate users' challenges in self-management of acute pain and their needs for these tools to help establish design opportunities.

In this paper, we approach this issue through an interview study with 14 patients who recently experienced postoperative acute pain. Through this study, we identified various factors related to the participants' challenges and needs that are associated with the major aspects of acute pain self-management; these aspects include the transition from hospital to home care, managing medication intake, handling the uniqueness of the pain experience, addressing emotional distress, navigating daily life challenges, working with caregivers, and receiving support from a broader social network. Our results revealed that the rapid changes in patients' postoperative pain experiences require both proactive and ``on-the-spot'' support. Moreover, the drastically different personal experiences of acute pain call for self-management support to focus on individual situations in context. Additionally, similar to chronic pain, self-management support for acute pain should concentrate on addressing patient motivation and taking into account their emotional states. These insights provided valuable design considerations of digital tools that can help alleviate challenges related to self-management of acute pain. 

\section{Background and related work}
Our study is situated in the related literature focused on (1) postoperative pain management, (2) technology support for self-management of health conditions, and (3) the design of technologies for pain management. We briefly review each body of literature in the following sections.

\subsection{Postoperative Pain Management}
Pain is considered a multidimensional experience comprised of sensory, emotional, and motivational dimensions~\cite{melzack1968sensory}, and understood within a biopsychosocial framework~\cite{Gatchel2007}. Its treatment and management thus require a multimodal, multidisciplinary approach. Pain is typically distinguished based on temporality, usually referred to as acute or chronic pain, when it lasts longer than three months~\cite{Treede2019}. Poorly managed acute pain is an important risk factor for chronicity~\cite{Page2020}. Among all the different types of acute pain, postoperative pain is very common.

Approximately 2.4 million surgeries are performed in Canada each year. Opioids play a central role in the management of acute postoperative pain, with 80\% of patients being prescribed opioids following surgeries~\cite{Wunsch2016}. This initially time-limited utilization becomes chronic for many patients~\cite{Page2020}. Long-term opioid therapy can lead to opioid addiction in about 6\% of the patients~\cite{Busse2017}. On the other hand, untreated postoperative pain can also have detrimental consequences. Acute pain remains poorly managed, with 11\% of patients reporting severe pain, and 37\% reporting moderate pain in the first day after surgery~\cite{Walker2016}. These statistics have been replicated in many countries~\cite{Gerbershagen2013, PQIP2021}. Postoperative pain management will typically involve pre-assessment and planning (when surgery is elective), assessment of pain, evaluation of treatment response, and treatment delivery~\cite{Small2020}. Typical treatment approaches will include the use of pharmaceuticals, such as paracetamol, non-steroidal anti-inflammatory drugs, opioids, ketamine, gabapentinoids, and other agents, administered using different modalities~\cite{Small2020}. The use of regional analgesic techniques is also recommended. Other approaches include education, psychological strategies, and physical therapies~\cite{Small2020}.

\subsection{Technology for Self-Management of Health Conditions}
Many previous works in HCI have focused on understanding and supporting self-management of various types of health conditions~\cite{Aarhus_Ballegaard_2010,Nunes2015, Burgess_Kaziunas_Jacobs_2022}. Most of these works focused on chronic conditions such as diabetes~\cite{Mamykina_Mynatt_Davidson_Greenblatt_2008}, abnormal blood pressure~\cite{Karnatak2023}, and migraine~\cite{Park_Chen_2015}, to name a few. Some have targeted mid to short-term conditions such as pregnancy~\cite{Gui_Chen_Kou_Pine_Chen_2017}, cancer~\cite{Eschler_Pratt_2017}, and injury~\cite{Stern_Howe_Njelesani_2023}. While the health conditions examined in those studies vary widely, our literature review revealed several overarching themes that cut across those studies.

First, in cases where self-management of health conditions starts when the patient is discharged from the hospital, the transition from hospital care to home care poses tremendous challenges. These challenges can be traced back to two fundamental differences between hospitals and homes~\cite{Aarhus_Ballegaard_2010}: (1) the physical environment and the material configuration shift from the care-centric hospital to the living-centric home and (2) the patient's role transforms from a passive recipient of care in the hospital to an active individual in charge, extending responsibilities beyond the health condition, when at home. To support the transition from hospital care to self-management, \citet{Pollack2016} argued that patients should be provided with support to develop three necessary ``self-management elements'': medical knowledge related to their health condition, resources and support to manage their health, and self-efficacy to learn and manage their condition. Built upon this framework, Tadas et al.~\cite{Tadas2023} designed and evaluated a system that uses patients' health-tracking and self-assessment data to help connect the clinical and non-clinical settings.

Second, social support is identified as a pivotal factor for self-management~\cite{Stern_Howe_Njelesani_2023}. The benefits of social support can come from patients' interactions in various social contexts. Interacting with \textit{health professionals} can promote patient's reflective thinking of their health conditions~\cite{Mamykina_Mynatt_Davidson_Greenblatt_2008} and augment patient agency~\cite{Ding2020}. Engaging the \textit{broader social circle}, including family, friends, and neighbors can provide practical and emotional support~\cite{Stern_Howe_Njelesani_2023} and help patients gain social recognition and acknowledgement~\cite{Park_Chen_2015}. Equally important, receiving support from \textit{peers with similar health conditions} can help alleviate isolation and self-awareness~\cite{Eschler_Pratt_2017}; additionally, individuals can gain advice, reassurance, and emotional support through these interactions~\cite{Gui_Chen_Kou_Pine_Chen_2017}. Incorporating these three layers of social support was identified as a key to achieving successful and effective self-management~\cite{Stern_Howe_Njelesani_2023, Nikkhah_John_Yalamarti_Mueller_Miller_2022}.

Third, despite the significant potential of technologies in facilitating self-management of health conditions, issues with technology use such as learnability and trust could hinder their adoption. To explain this phenomenon,~\citet{Burgess_Kaziunas_Jacobs_2022} identified several ``care frictions'' that capture the discrepancy between the ideal but unrealistic patient behavior in health technology design and the practical challenges and concerns of patients in real life. For example, technology design often presumes that patients are motivated to read information relevant to their health and communicate with their healthcare team; in reality, however, patients may be already overwhelmed with information or have personal concerns when communicating with the clinicians~\cite{Burgess_Kaziunas_Jacobs_2022}. Patient trust in self-management technology is also a complex issue. A study with low-resource communities in India identified that trust in technology is influenced by many social-cultural factors, including trust in care providers and broader social circles~\cite{Karnatak2023}.

Our work is built upon this body of literature and focuses on the real-world challenges and needs of patients who experience postoperative pain. This type of health condition happens fast and thus may provide unique implications to transition to home care, social support, and friction in technology use.

\subsection{Design of Technologies for Self-Management of Pain}
User-centered design for chronic pain management applications has been well-researched in the literature. Among the existing efforts, supporting self-assessment of pain is often considered the first step toward effective self-management~\cite{10.1145/3025453.3025832, Adams2018Keppi}. Because pain is a highly personal and subjective experience~\cite{Koyama_2005}, self-assessment of chronic pain needs to consider individual differences and incorporate their contexts. Previous research has established that patients' preferences on the ways to report their pain experience using mobile interfaces are ``strong and idiosyncratic — and not always unanimous''~\cite{10.1145/3025453.3025832}, highlighting the need to provide personalized pain assessment measures. In pain assessment techniques that leverage automated technologies, such as a tangible user interface that uses the squishing motion to report pain~\cite{Adams2018Keppi} and a smartwatch technology that detects fist clenching gestures~\cite{Ramprasad2021}, the importance of accommodating personal preferences in the physical and aesthetic design is also often stressed~\cite{Adams2018Keppi}.

Motivating chronic pain patients to engage in supportive activities is another important and widely recognized issue. Different from clinical rehabilitation sessions, self-management of chronic pain needs to engage with patients' daily lives~\cite{Singh2017}. Focusing on encouraging physical activities,~\citet{Singh2014} found that the design of motivational technologies for chronic pain patients needs to consider their emotional states, such as anxiety that may result in avoidance. Taking this into account, the authors designed an aural feedback mechanism to support patients in building confidence in movements that they fear because of expected pain. Previous research also identified that using immersive technologies such as VR can motivate patients to engage in activities that aid pain alleviation~\cite{Wang2022} and encourage and teach them meditation techniques to reduce pain~\cite{Gromala2015}.

On the other hand, however, there is very limited previous work that targeted the design of technologies for postoperative acute pain management. We only identified a few sporadic efforts in our literature review. Koumpouros~\cite{10.1145/3453892.3461350} described an approach for designing a mobile application to support ``management, reporting, and treatment effectiveness monitoring'' for patients experiencing pain; although the app is claimed to focus on general pain management, the design emphasized chronic pain a lot more than acute pain. Lalloo et al.~\cite{Lalloo_2017} also conducted a review of commercially available health apps that can potentially help postoperative pain self-management; their results highlighted that (1) none of the commercial tools have provided comprehensive pain self-management content and (2) none of them have involved end users (i.e., postoperative patients) in their design and development. To address some of these issues, Birnie et al.~\cite{Birnie2019} designed a mobile app for postoperative pain management through a user-centered approach; however, their app only targets children and adolescents. Results like these indicate the need to investigate the user requirements and the design of tools to help patients with their self-management journey after surgery. We provide some important initial efforts to address this gap.

\section{Methods}
We conducted semi-structured interviews between June 2021 and August 2022, with 14 participants from Quebec, Canada to develop an understanding of how individuals experience and manage their postoperative pain. The study is approved by the research ethics boards of the authors' institutions. The rest of this section discusses the procedures we followed to collect and analyze our data.

\subsection{Participants}
We recruited, through flyers posted within the hospital and on social media, 14 participants who underwent major surgery in the previous 12 months that required a prescription of opioid analgesics to manage acute postoperative pain. Table~\ref{tab:participants} summarizes their characteristics. The average age of the participants was 45.6 ($SD=16.2$, ranging from 19 to 77); 80\% were female and 20\% were male. Out of these 14 participants, seven had orthopedic surgery, two gynecology surgery, one neurosurgery, and one digestive surgery; three did not specify their surgery types. On average, the participants completed the interview 20 weeks after their surgery (ranging from 1 to 52 weeks). Among the participants, eight had never taken opioids before their surgery, while the rest did in a regular manner for their health condition. Ten participants described their pain after surgery as severe, while four described it as moderate.

\begin{table*}[t]
\small
\caption{Characteristics of interview participants}
\begin{tabular}{lllll}
\hline
\textbf{ID} & \textbf{Gender} & \textbf{Highest level of education} & \textbf{Employment status} & \textbf{Weeks after surgery} \\
\hline
P01 & Female & University & Retired & 2 \\
P02 & Female & University & Sick or maternity/paternity leave & 5 \\
P03 & Female & University & Other & 26 \\
P04 & Female & Technical college & Disability for a reason other than pain & 52 \\
P05 & Male & Secondary School & Sick or maternity/paternity leave & 4 \\
P06 & Female & University & Full time job & 31 \\
P07 & Female & University & Full time job & 25 \\
P08 & Female & University & Full time job & 40 \\
P09 & Female & University & Other & 13 \\
P10 & Female & University & Full time job & 2 \\
P11 & Male & Secondary School & Disability associated with pain & 3 \\
P12 & Female & University & Retired & 45 \\
P13 & Female & Technical College & Full time job & 31 \\
P14 & Male & Technical College & Full time job & 1\\
\hline
\end{tabular}
\label{tab:participants}
\end{table*}

\subsection{Interview Procedure}
Each participant was compensated with \$30 CAD for participating in the interview. Each semi-structured interview session lasted about 60 minutes. All interviews were conducted over Zoom. During the interviews, we asked the participants questions about their experiences, challenges, needs, and strategies used in postoperative pain management. For example, we asked them to describe their pain experience after surgery, enumerate the most challenging aspects of pain management, reflect on the techniques, tools, or strategies that helped them manage pain, and tell successful and unsuccessful stories when managing the pain. We also asked questions related to their general views on technology use for pain management, including the type of information they need for better pain management and the essential features they wish to see in an ideal tool. The interviews were conducted in either French or English, based on the choice of the participants.

\subsection{Data Collection and Analysis}
Interviews were video recorded and then fully transcribed; the interviews conducted in French were also translated into English. We conducted a thematic analysis~\cite{Vaismoradi2013} on the interview transcripts using Atlas.ti\footnote{https://atlasti.com}. Three researchers were involved in the iterative, inductive data analysis through independent coding and extended discussions. Themes were identified along the major aspects of self-management of postoperative acute pain, focusing on participants' challenges and needs.

\subsection{Study Context}
\label{sec:context}
There are two factors in the context of our study that are important to understand. First of all, our study is conducted in Quebec, Canada. Thus our participants' experiences are contextualized in the \textbf{Quebec health and social service system}. In Quebec, the health and social service system operates at two main levels: (1) the Ministry of Health and Social Services is responsible for creating, monitoring, regulating, and coordinating the key strategic directions and (2) the health and social service centers are responsible for directly providing primary care and some specialized services to different regions of the province~\cite{MSSS2015}. The majority of the physicians are paid by the public health insurance system and their services are offered for free to eligible persons covered by the Quebec Health Insurance Plan. However, while pharmacological treatments are mostly covered through public or private insurance, most physical (e.g., physiotherapy) and psychological treatments are private and limited reimbursement through private insurance for a limited number of individuals. At the same time, patients can experience long wait times before receiving the services~\cite{Desmeules2010, Delaurier2013}. Our participants are aware of the usually long wait times and expressed their frustration, describing their pain management journey as  ``\textit{do it yourself}'' (P02) and ``\textit{sink or swim}'' (P10). 

Second, our interviews were conducted between June 2021 and August 2022. This time span covered some of the peak periods of the \textbf{COVID-19 pandemic} in Canada. While in the study and the analysis, we did not directly focus on postoperative pain management during the pandemic, this is an important context to consider when interpreting some of our results. Our participants also frequently mentioned the impact of COVID-19 on their postoperative care. For example, P01 said, ``\textit{It is my first surgery experience and I didn't find it the most wonderful on earth -- the COVID played a big part I think.}'' Overall, the COVID-19 pandemic put constraints on the surgery procedure (e.g., family members not allowed in the hospital), increased burdens of the public health system, and rendered even fewer resources to the patients postoperatively, aggravating the frustration and anxiety of our participants. For example, P04 said, ``\textit{We are in times of COVID, and unfortunately, the nurses, the attendants are super busy. They are not always available when you need them. So I found it really hard.}''

\subsection{Researcher Positionality}
We acknowledge that our positionality as researchers shaped how we collected and analyzed the data. Because the topic of this study is related to postoperative patients' experiences and their perception of digital tools, we think it is especially important to explain the background and expertise of the research team members. In the research team who authored this paper, two are experienced in the field of human-computer interaction (HCI), with a specific focus on the human-centered design of emerging and smart technologies; one author is a researcher and clinical health psychologist who has expertise in the transition from acute to chronic pain and opioid use in the context of surgery; two authors are patients who have experienced first-hand post-operative pain and/or prolonged hospitalization associated with surgical complications. Particularly, the two patient authors~\cite{Oliver2022} were involved throughout the research project to provide their input and help us better empathize with our participants, taking part in activities such as (1) framing the research questions, (2) developing and refining the interview guide, (3) conducting the interviews, (4) analyzing the data and interpreting the results, and (5) revising the manuscript. Each interview was conducted jointly with two authors -- one HCI researcher and one patient author or health researcher; this composition was used to ensure that both technological and health-related aspects were sufficiently discussed during the interview. The data analysis was led by the HCI researchers, with inputs from the other authors, to focus on implications towards technology design.
\section{Results}
In the following sections, we report the themes we identified that are associated with the major aspects of self-management of postoperative acute pain. Because pain self-management is a holistic process, these aspects may overlap with each other.

\subsection{Transitioning From Hospital to Home Care}
While several participants faced postoperative complications that extended their hospital stay and aggravated their pain experience, the majority stated that their post-anesthesia pain was controlled and tolerable. However, many participants shared a feeling of rushed hospital leaves; e.g., ``\textit{they are in a hurry so that we go home}'' (P07). This was sometimes due to the over-capacity state of the hospital, especially during COVID-19. Whatever the reason, the \textbf{rushed hospital leaves caused great frustrations and uncertainties} for our participants. For example, P10 described vividly their experiences after discharge: ``\textit{I got out of the hospital... and it's `sink or swim.' They put you in the water. You don't swim, but you organize yourself to get yourself at the surface, and you hope you'll succeed. It's a bit like that.}''

Surgery is often a rare life event and can happen because of an accident. Thus, participants mentioned that they \textbf{needed assistance to quickly establish expectations, even before the surgery}. Such assistance is important for the participants as it helps them to prepare for the interaction with the medical team and to better digest and contextualize the information received. For example, P09 said, ``\textit{I think for me, probably is to have a checklist, to ask the surgeon before or after the surgery. ... I didn't know what to expect. So, if I have a list of how to cope with anesthesia, how to cope with the pain, and what kind of side effects after surgery, I need to be aware of or pay attention to... that would be helpful.}'' 

Our participants also mentioned that \textbf{once transitioned at home, new health conditions may emerge}. For example, P03 said, ``\textit{When I got home, there was another kind of pain that happened. You know, the same pain persisted... I was able to understand that, but there were other kinds of pain that happened that I didn't expect, like electric shocks, and I really didn't know how to deal with it.}'' Getting answers to the emerging questions was difficult, if at all possible, causing stress and frustration in pain self-management. 

\subsection{Managing Medication Intake and Its Side Effects}
One of the biggest challenges after transitioning to home care was related to dealing with the overwhelming responsibility when self-managing postoperative medications. For example, P10 mentioned: ``\textit{Painkiller, it was something to manage... I started with a dose that was administered in the hospital when I had pain, which was obviously supervised by the doctors... Now I have to manage it myself in proportion to the time when I had the pain. Well, that was a big step.}''

Participants discussed a common \textbf{challenge in understanding medication, including when and how to take it, as well as its side effects}. This is sometimes due to the lack of information or explanation at discharge. For example, P07 mentioned: ``\textit{The doctor prescribed it [opioid-based pain killer] to me twice a day, and then he gave me, say, thirty tablets... [Now I understand that] we take them when we have pain, and not to make a habit of taking them every 12 hours... But I wish there were education that is done on that side.}'' 
With insufficient knowledge, participants sometimes reduced medication intake or even stopped it altogether because of the fear of side effects (e.g., dependence). For example, P09 mentioned, ``\textit{When I looked at the information, I don't know whether... Even for Tylenol, I also worry I am taking too much. I don't know. And is it going to cause problems?... I don't believe that there are no side effects for any medicine... So, I still don't want to continuously rely on that.}''

In addition to the knowledge gap, participants also discussed the \textbf{difficulty to keep track of their medication intake as it becomes part of their new postoperative routine}. For example, P05 said, ``\textit{I never knew when to actually take [my medication], and I really don't have a good memory. So sometimes I forget when it is already an hour later or something.}''
P13 also discussed, ``\textit{I often get lost when I take my medication. I don't have the habit of writing it down, but sometimes, I don't know what time I took my medication... and I don't know if... [I can take it again or not].}'' Participants also discussed the usefulness of keeping track of the effects of the medication, such as its pain impact and side effects. Regarding this,  P03 mentioned, ``\textit{I kept a note... I wrote I took such medicine at such time, in such quantity... Then you can monitor your pain, you know, like at that time I was a 2 or I was an 8... And also, perhaps to track it or see `oops, on such a day, there was a peak'...}''

Moreover, the intake schedule of many acute pain medications should be based on necessity and pain tolerance, in addition to timing (many with a minimum interval). Thus, participants discussed \textbf{an intricate problem of balancing medication intake and non-drug pain relief methods}. For example, P10 discussed elaborately how they managed opioid intake: ``\textit{Because when you're in pain, you have to manage it like that... You hang on till the moment when you are able to, when I give myself permission to take [the medication]. But on the other hand, I was so proud when I went straight, and I didn't need it. That day, I took one shot less. It's wonderful. I would be really proud of myself.}'' Traditional medication reminders may not be helpful, or even cause harm in this context, as P10 continued to explain: ``\textit{I wouldn't even want an alarm that reminds me that it's okay to take ... for example, Dilaudid. If I'm passed four hours or five hours, I have the pain to say to myself `Okay, I need it.' But I wouldn't want an alarm to tell me `Well, after three hours, you now can take some.'}'' To manage pain without medication, many participants adopted techniques such as meditation, music, breathing, and physical activities to help them through their pain episodes.

\subsection{Managing the Uniqueness of the Pain Experience}
\label{res:uniqueness}
The participants frequently referred to their pain experience as ``unique'' and ``subjective''. Because of this, participants frequently discussed \textbf{the challenge of self-assessing the ``normalness'' of their pain experience}. For example, P01 discussed, ``\textit{I was a bit lost on `okay, I have an exercise to do. They say I shouldn't have pain, but I have pain all the time.' So, I don't know if I have to continue it or not... I was like `Am I being a bit of a sissy', or `Is this normal at all?'}'' The uniqueness of the pain experience also made it counterproductive to compare with other patients who had similar conditions. What worked for someone does not automatically mean it would work in another case; what is experienced by someone does not mean it is ``normal.'' For example, P01 mentioned: ``\textit{Everyone has their experiences, and we always hear a lot of things... I had a friend who had exactly the same surgery two months before me, except that she seemed to be doing super, super well. And it's for sure that suddenly, we didn't have the same evolution at all... I'm normal, am I not? Is there something wrong?}'' So the supposedly comforting practice of sharing and discussing their experience may become ``\textit{a source of anxiety}'' (P01).

The complex pain experience also made it \textbf{difficult to assess pain in isolation, using quantitative metrics}. A typical assessment of pain is a 10-level rating scale. However, our participants noted how unnatural it feels to assign a number to pain, indicating the rating system to be insufficient to capture how they feel. They believed that pain can be described differently, qualitatively. For example, P10 said, ``\textit{I found it a little hard to put a number on it.}'' P02 also said, ``\textit{What does 1 to 10 mean? ... There are times when the pain leads you to scream that you can no longer control yourself, there are others... If you can describe the pain more specifically, it would give you a picture that would be more precise about your pain.}'' Moreover, participants mentioned different symptoms associated with the surgery that can be used as a proxy to indicate pain. For example, P06 talked about the need ``\textit{to validate if the swelling is bigger, does that change...}'' P07 also mentioned the importance of tracking inflammation at the wound, ``\textit{because definitely, when there is more inflammation, there is more pain.}'' Participants also discussed the importance to consider the context and situation that the patient is in when monitoring pain. For example, pain during exercise is not necessarily a bad thing, since ``\textit{the principle of doing physical management, moving, with pain management, in any case, really goes hand in hand}'' (P10).

The pain experience of individual patients can also develop and vary over time, ranging from mild and controllable to intense and severe levels. Our participants discussed \textbf{the importance of receiving support when pain is in extreme intensity}. They explained that those pain episodes often came fast and in severe cases, they simply forgot what to do when in pain. So, our participants discussed the need to have tools that can help them detect such critical situations and redirect them to adequate help. For example, P06 expressed the need to have a ``panic button'' that accelerates their connection to medical support: ``\textit{To have the possibility of when things are not going well -- to have something at the end of the line that will help us, fast...}''

\subsection{Managing Emotional Distress}
Our participants shared \textbf{a mixture of emotions of being overwhelmed, helpless, depressed, and anxious}. For example, P04 said, ``\textit{You know, emotionally and all that, it was very hard. It was a lot.}'' P06 also mentioned: ``\textit{I was like, `Well. Nothing works that well for me, so I don't know what it will look like over time.'}'' These types of emotional distress were mainly triggered by the occurrences of unexpected experiences and the differences between their pain expectations and the actual pain experienced. Overall, there was a fear of the unknown. P03 summarized this: ``\textit{Maybe it comes down to `you don't know what you don't know.' ... For me, in all my experience, all the uncertainties or all the things we don't know, when it hits you, it seems like it's even more tiring...}''

\textbf{The intensity of the emotions varied based on the pain experiences of the patients at different times.} For example, P06 described how their emotional states changed from day to day depending on the level of pain they had: ``\textit{I wasn't irritable or anything in general. But you know, there were nights that I found it really difficult -- like, if I could give a little bit of my pain to someone else, I would have done it (laughs). I would have given quite a bit, even (laughs). I would have made a small donation.}'' P06 also said: ``\textit{The pain made me numb. Maybe also with the medication, but I was in so much pain that I didn't feel like doing things. I didn't recognize myself, not at all.}''

At the same time, participants also discussed that \textbf{negative emotions can result in a higher intensity of pain, making it a vicious circle}. For example, P03 said, ``\textit{So for me, [the pain] was difficult, and worse, not knowing what to do to relieve it. So that, well, more than you are in pain, unfortunately, the reflex is that we tense up, well it makes it worse...}'' P13 also mentioned, ``\textit{I'm going to have an anxiety attack or whatever, well the pain will increase too.}'' Participants discussed that management of emotion is an important part of self-management of postoperative pain, as P13 continued: ``\textit{The fact is that it's also about trying to manage the anxiety. Because when I have a pain attack, there's always an anxiety attack that comes with that. But with the anxiety attack, the pain just increases.}'' Participants described various strategies to deal with their emotions like seeking reassurance from people around them, taking time for self-care, sleeping when overwhelmed with emotions, and hobbies like sports or music that kept their minds busy.

\subsection{Managing Daily Life Challenges}
The majority of our participants mentioned their \textbf{struggles to keep up with everyday chores and duties}. Challenging tasks were as simple as meal-prepping, housekeeping, and personal hygiene obligations such as bathing, dressing, grooming, and toileting, among other routine tasks. As P06 put, ``\textit{it's the loss of autonomy.}''

\textbf{The level of daily-life challenges was dependent on various factors}, including participants' personal obligations, temporal context, and the type of support they have. First, \textit{personal obligations} such as taking care of pets (P01), moving houses (P02), family duties (P09), etc., affected how many daily-life challenges participants have to deal with. They often felt obligated to complete these tasks, even when they were pushing their physical and emotional limits. Second, \textit{temporal contexts} also played a role. For example, P09 described their increased challenges during nighttime: ``\textit{The pain was manageable during the day time when I'm interacting with others -- it basically alters the mind. And during the evening when I went to sleep, it was quite challenging. In the middle of the night. I often get up and then the whole pain, and also just the overall discomfort of the cast and everything just made it hard to fall back to sleep.}'' The harshness of the challenges also varied based on \textit{the type of support} the participants had at home. For example, P01 said, ``\textit{What didn't help me either was that I had no help at home, so I had to do everything on my own. So maybe I was moving too much too.}''

On the other hand, participants mentioned that although challenging, \textbf{succeeding in accomplishing mundane activities provides a sense of achievement} that can serve as an important encouragement during their recovery process. For example, P07 said, ``\textit{I find that it is more rewarding if we are able to complete tasks -- whether it is washing the dishes, doing the laundry, doing a little housework, preparing a meal... So sometimes, just targeting more functional tasks in the house helps, since it's more general.}'' Ability to resume daily activities also serves as a motivator to engage in rehabilitating exercises, as P06 said: ``\textit{I am motivated to be able to, first, walk again, second, ride a bike, play sports, and maybe one day run, ski... That was what motivated me.}'' 

\subsection{Working With Caregivers}
Our participants emphasized \textbf{the crucial role their at-home caregivers played in their postoperative journeys}. They were often the people who accompanied them from the hospital, received the prescriptions, helped them understand the situation, managed their medication intakes, pinpointed red flags in their conditions, helped them take injections and change bandages, etc. E.g., P04 explained, ``\textit{At home, it was easy, because my mother was there. Okay, Mom has always been there since the beginning of my illness, and you know if I am not well, she will see it and she [will help]... And it is she who manages the Fentanyl patches too. Makes her change them and all that, every three days.}'' While essential to the recovery of the patients, these responsibilities can be onerous to the caregivers. As P10 explained: ``\textit{And my spouse who was a caregiver... He is the most important. I think they are the ones who suffer the most, more than the patient ... Because there's nothing that I consumed without passing it through him.}'' 

Our participants also discussed that, since their caregivers often do not have sufficient knowledge and experience, \textbf{caregivers often struggled between a strong desire to be of assistance and being clueless about how to do it}. For example, P04 said, ``\textit{But instinctively, [my mom] saw when it was a little infected, and when it was this, when it was that. But, I mean, she did everything on instinct. She didn't have a class. No one told her what to do. And you know, I think that must have stressed her out somewhere. That's a big load for someone.}'' Because of the lack of adequate knowledge and experience, the caregivers might have inaccurate assumptions. Addressing these assumptions and maintaining a positive relationship with the caregivers became a challenge and a new source of stress for some of our participants. For example, P01 explained: ``\textit{People have a representation in their heads of what we can do, what we can't do, even if they don't know anything about it. And sometimes it adds to the mental load that we already have to manage our pain and our post-surgery. Worse, you have to explain again `Yes, no, I can't do that, but I can do that.' But they, in their mind, if [they think] we can't do that, we shouldn't even try. And there, it is like an additional challenge.}''

\subsection{Receiving Support From a Broader Social Network}
Despite the uniqueness of pain (see Section~\ref{res:uniqueness}), participants mentioned that \textbf{they wanted to learn how other people with similar conditions experienced their pain}, their recovery timelines, their medications and the side effects, along with helpful tips to battle pain. About this, P05 said, ``\textit{After the operation, they say `go home, take your medicine.' And no information on how you will feel or how the body will react. So it's fun to know about other people's experiences.}'' P10 also expressed their enthusiasm towards having an experience-sharing group: ``\textit{If I can volunteer, I can contribute to programs like this... `Oh my God.' I'll be the first to raise my hand.}'' 

Since pain is also an emotional experience, some of our participants, who were particularly lonely with no support at home, \textbf{wished to have social support that can help them emotionally}. Regarding this, P03 said: ``\textit{But you know, I live alone. The fact that having someone ... who is there to be able to respond to this kind of situation or to try to calm my nerves -- that would have certainly contributed to having the same effect, or even more so, to soothe.}''

On the other hand, participants also discussed \textbf{potential issues with experience sharing}. They explained that they are concerned about negative messages that they may receive from social groups. For example, P05 said: ``\textit{The `counter-side' is that sometimes it's not that positive and it's exaggerated. It's their situation, but sometimes it's extremely negative. In that case, it just causes even more anxiety.}'' Participants were also afraid that social groups may propagate inaccurate or incomplete information. For example, P07 said, ``\textit{Each person comes with their share of comorbidities... Well to say `Ah, me, it went well, and you know, after a week, I was back on the peak.' Maybe it's true for a young person who was in good health, who has no family, who has no young children to take care of at home. But for a single mom, she may not have the same situation. So I don't know if it's a double-edged sword...}''
\section{Discussion}
Through our qualitative analysis, we identified various challenges and needs of postoperative patients regarding pain management, which have implications for the corresponding technology use. In the following sections, we synthesize our results, examine them in the context of related literature, and discuss design implications for postoperative pain management technology. 

\subsection{Rapid Changes Require Both Proactive and ``On-The-Spot’’ Support}
Different from many chronic health conditions, patients undergoing postoperative acute pain frequently encounter swift changes in their health. These changes often start at the surgery and hospital discharge. Previous works have established that transitioning from hospital care to home care posed considerable challenges for the self-management of various health conditions~\cite{Aarhus_Ballegaard_2010}. Our work confirmed and emphasized this as a major obstacle for postoperative patients. Many of our participants noted a lack of sufficient information regarding the surgery, strategies for health and pain management, and the proper approach to medication intake. They also expressed worry and anxiety about getting lost and being unsure of how to take care of themselves. In other words, they lack all the key elements (i.e., knowledge, resources, and self-efficacy) for self-management of pain~\cite{Pollack2016}.

During the progression of pain after discharge, quick changes may also occur. This is a factor that is less discussed in the literature that is mostly focused on chronic health conditions. Our participants frequently discussed that after transitioning at home, new health conditions, including new types of pain, may surface rapidly. In certain circumstances, situations can quickly deteriorate. Intense pain may strike without precaution or happen at unexpected or inconvenient times (e.g., at night). In those situations, participants often felt helpless and desired to have immediate support for pain alleviation.

The moments such as hospital discharge and intense pain episodes are the ones in which participants require support most urgently. However, those are also the moments in which the patients are the most vulnerable. That is, they were usually in an inferior health condition (e.g., pain, dizziness, vomiting after the surgery) that prevented them from processing and remembering the needed information. As a result, providing information or support only at those moments is usually counterproductive. At the same time, during the recovery path, various factors take a toll on the participants' mental and cognitive states. They are already overwhelmed with all the variables they have to manage among pain, medications, chores, etc., making it hard for them to digest and remember complicated guidelines and advice provided to them.

Digital tools for postoperative pain management thus need to carefully consider the timing and the manner of information and support delivery. Depending on the patient's conditions and health literacy, for example, dumping a large amount of information to ``prepare'' the patients may not be a good idea. Instead, bite-sized information or suggestions suitable for the current recovery stage could be more appropriate. Illustrations, visualizations, and short animations are generally preferred over lengthy textual explanations. At the same time, when unexpected events happen (e.g., severe pain in the middle of the night), patients should be able to get quick support with minimal interaction.

\subsection{Self-Management Support Should Focus On Individual Situations in Context}
A recurring theme that emerged from our analysis is that postoperative acute pain is a personal experience that is unique to each individual. This is related to pain assessment, a topic that is repeatedly discussed in the chronic pain management literature (e.g.,~\cite{10.1145/3025453.3025832, Adams2018Keppi}). Similar to \citet{10.1145/3025453.3025832}, our participants find it difficult to use the traditional 10-level rating scale for pain assessment. More importantly, our participants emphasized that their pain experience is not linear (e.g., low or high levels), but something much more complex, perhaps multi-dimensional, associated with the type of pain, the part of the body where the pain happens, and the symptoms that co-occur with it. The uniqueness of pain has implications and affects various aspects of postoperative acute pain self-management, such as social support and medication management.

Social support is often considered a positive aspect during self-management of health conditions~\cite{Stern_Howe_Njelesani_2023}. Our participants, however, painted a more complex picture due to the uniqueness of their acute pain experience. On the one hand, echoing the literature, they valued the support provided by their caregivers and peers, considering the importance of both practical and emotional support from their social network~\cite{Stern_Howe_Njelesani_2023, Eschler_Pratt_2017}. On the other hand, however, participants felt that it is difficult to compare their experience with others or communicate their pain to others who are not experiencing it. The social network of peers, which is often considered a source of advice and reassurance~\cite{Gui_Chen_Kou_Pine_Chen_2017}, becomes a ``source of anxiety'' for our participants, making them frequently question the ``normalness'' of their pain. Caregivers, despite their well-meaning intentions, are frequently overwhelmed and struggle to understand the patient's needs, rendering their support at times irrelevant or unhelpful in reality. This complexity of social support needs to be incorporated into the design of technology for postoperative pain management.

Due to the individual differences in the pain experience, medication in acute pain self-management is particularly a tricky issue. This is because many acute pain medications, especially opioid analgesics, have severe side effects, such as addiction. Patients were often only given very general guidelines on how to take the pain relief medications. They then have to decide by themselves based on their unique pain experiences, pain tolerance, and personal strategies for non-drug pain relief. Participants frequently discussed their reluctance to take pain medicines and shared personal  techniques, like deep breathing and listening to music, that proved effective for them in alleviating pain. A high-level guideline like ``take this every four hours or more when pain is severe'' does not help very much according to our participants. A more personalized advice, or even better, an approach helping patients discover a strategy that fits their individual characteristics, is considered more useful.

Considering all the above factors, tools should provide personalized and dynamic support that can adapt to the particularities of individual patient's health conditions in their immediate context. This requirement also applies to assistance in managing social support and medication. For example, when it comes to peer support, the matching of ``similar'' patients should be carefully considered, since the definition of ``similarity'' can depend on various factors, and not all comparison to ``similar'' cases is helpful. Educating caregivers and supporting them to better understand and assist the pain management process of individuals is also useful. Similarly, medication management support should incorporate personal preferences in non-drug pain relief strategies and avoid periodical reminders.

\subsection{Managing Emotion and Motivation Has Important Repercussions}
Previous research has identified that technologies that support self-management of pain should focus on motivating the patients and consider their emotional states~\cite{Singh2014, Wang2022}. Our results confirmed the importance of these factors in acute pain self-management. Interestingly, our participants discussed a vicious circle formed by pain and negative emotions -- increased pain would lead to anxiety and helplessness, which in turn, would result in more severe pain. Thus, strategies that help participants steer away or jump out of this vicious circle will have important impacts.

Participants also discussed various ways and specific occasions that helped elevate their emotional state. Social support from caregivers and peers is important, albeit with its own risks (see the discussion above). Seemingly trivial things, such as accomplishing mundane activities, can usually provide an important sense of achievement, fostering a heightened positive emotion and motivation. Leveling the difference between the patient's expectations and the actual experience can also pave the way for dispelling doubt and anxiety.

Thus, monitoring emotion and promoting motivation should be an important consideration in tools for supporting postoperative pain management. Tools should track the patient's recovery progress, help them establish realistic goals, and facilitate their own autonomy and self-efficacy in pain management. Related, tools should focus on supporting patients in making the right decisions themselves, rather than imposing a suggestion or recommendation. The patients should be able to experiment with different strategies themselves and make choices that align with their priorities and values, rather than being dictated to do things.

\subsection{Limitations and Future Work}
There are several limitations of this study that can benefit from future work. First, as we described in Section~\ref{sec:context}, the study was done during the COVID-19 pandemic, in Quebec, Canada. Thus, our results and recommendations need to be considered under this context and its applicability to other contexts needs to be explored. Second, we used the interview method to understand the participants' challenges, needs, and desires. While this method provided rich results reflecting participants' values and perceptions, other methods such as an observational study or a diary study can help better understand participants' actions or experiences over time. Third, our participants had a diverse profile in terms of the types of surgery they underwent and the time passed since their surgery. While this diversity allowed us to identify the most prominent challenges and needs in pain management across a wide range of postoperative patients, we did not analyze the differences along these patient profile dimensions and leave this to future investigations. Finally, in this study, we focused on understanding the need for tools for postoperative pain management from the patients' perspectives and did not analyze the existing tools. While the current tools for pain management almost exclusively focused on chronic pain (e.g., Manage My Pain\footnote{https://managinglife.com} and Curable\footnote{https://www.curablehealth.com}) and postoperative pain has its unique trajectory (e.g., supposed to improve quickly over time) and challenges (e.g., severe side effects of medication), analyzing the existing pain management tools may provide important insights and should be done in future work.

\section{Conclusion}
Through an interview study with 14 postoperative patients, we focused on understanding their pain self-management experiences and identifying their requirements for digital tool support. Our analysis identified various factors associated with the major aspects of acute pain self-management. Together, these results indicated that, first, digital tools designed for postoperative pain management need to carefully consider the timing and the manner of delivering information and support to adapt to the rapid changes that may occur during the patient's pain experience. Additionally, these tools should offer personalized and dynamic support that can adjust to the specific conditions of individual patients within their immediate context. Moreover, monitoring emotion and promoting motivation should be an important consideration when designing tools for supporting postoperative pain management. Overall, our study provided valuable knowledge to address the less-investigated but highly-needed problem of technology design for the self-management of acute pain. Our efforts can inform other studies focused on the design of supportive technologies for similar health conditions.

\begin{acks}
We thank our participants for their time and thoughtful answers. This research is partially supported by the Fonds de recherche du Québec (\grantnum{FRQ}{2021-AUDC-284951}) and the Canada Research Chairs program (\grantnum{CRC2}{CRC-2021-00076}).
\end{acks}

\bibliographystyle{ACM-Reference-Format}
\bibliography{references}

\end{document}